\documentclass[aps,twocolumn,prb,eqsecnum,showpacs]{revtex4}
\usepackage{amsmath,amssymb}
\usepackage{graphicx}
\begin{document}
\draft
\preprint{17 December 2009}
\title{Ground-state properties of a Peierls-Hubbard triangular prism}
\author{Shoji Yamamoto$^1$, Jun Ohara$^2$ and Masa-aki Ozaki$^3$}
\address{$^1$Department of Physics, Hokkaido University,
         Sapporo 060-0810, Japan}
\address{$^2$Division of Physics, Yokohama National University,
         Yokohama 240-8501, Japan}
\address{$^3$Uji 611-0002, Japan}
\date{17 December 2009}
\begin{abstract}
Motivated by recent chemical attempts at assembling halogen-bridged
transition-metal complexes within a nanotube, we model and characterize a
platinum-halide triangular prism in terms of a Peierls-Hubbard
Hamiltonian.
Based on a group-theoretical argument, we reveal a variety of valence
arrangements, including heterogeneous or partially metallic
charge-density-wave states.
Quantum and thermal phase competitions are numerically demonstrated with
particular emphasis on novel insulator-to-metal and insulator-to-insulator
transitions under doping, the former of which is of the first order, while
the latter of which is of the second order.
\end{abstract}
\maketitle

\section{Introduction}

   The success of patterning a graphite sheet into a cylinder
\cite{I56,E220,K878} stimulated the public interest in the geometric
tunability of electronic properties.
Carbon nanotubes indeed vary from metals to semiconductors with the inside
diameter and the chiral angle. \cite{C4266}
Their nature is fairly describable within the $\pi$-electron
nearest-neighbor tight-binding model \cite{S2204,S1804} and is relatively
insensitive to Coulomb interactions.
A tubed vanadium oxide, \cite{M676} Na$_2$V$_3$O$_7$, consists of strongly
correlated $d$ electrons, but they are completely localized.
Another transition-metal-based nanotubular compound, \cite{S1580}
[Cl(CuCl$_2$tachH)$_3$]$X_2$
($X=\,$Cl, Br;
 tach$\,=cis,trans$-$1,3,5$-triamino-cyclohexane$
      \,=\,$C$_{18}$H$_{45}$N$_9$),
also behaves as a Heisenberg magnet.
In such circumstances, a platinum-iodide quadratic-prism compound,
\cite{Ounpub}
[Pt(en)(bpy)I]$_4$(NO$_3$)$_8$
(en$\,=\,$ethylendiamine$\,=
 \,$C$_2$H$_8$N$_2$;
 bpy$\,=4,4'$-bipyridyl$\,=\,$C$_{10}$H$_8$N$_2$),
featured by competing electron-electron and electron-lattice interactions,
has sparked a brandnew interest \cite{O17006} in lattice electron
nanostructures.

   Quasi-one-dimensional transition-metal ($M$) complexes with bridging
halogens ($X$) possess unique optoelectronic properties,
\cite{M5758,M5763,G6408,W6435} which are widely variable according to
the constituent metals, halogens, ligand molecules and counter ions.
\cite{O2023}
Platinum-halide single-chain compounds such as Wolffram's red salt
\cite{C475}
[Pt(ea)$_4$Cl]Cl$_2\cdot$2H$_2$O (ea$\,=\,$ethylamine$\,=\,$C$_2$H$_7$N)
have a Peierls-distorted mixed-valent ground state, whereas their nickel
analogs \cite{T4261,T2341} consist of Mott-insulating monovalent regular
chains.
In between are palladium halides, whose ground states can be tuned
optically \cite{I241102,M123701} and
electrochemically. \cite{M7699,M035204}
Metal binucleation considerably stimulates the electronic activity.
\cite{Y125124,K2163}
A diplatinum-iodide chain compound, \cite{K4420}
$R_4$[Pt$_2$(pop)$_4$I]
[pop$\,=\,$diphosphonate$\,=\,$P$_2$O$_5$H$_2$;
 $R=\,$(C$_2$H$_5$)$_2$NH$_2$],
exhibits photo- and pressure-induced phase transitions,
\cite{S1405,Y140102,Y1489,M046401,Y075113} while its analog without
any counter ion, \cite{B444}
Pt$_2$(dta)$_4$I
(dta$\,=\,$dithioacetate$\,=\,$CH$_3$CS$_2$),
is of metallic conduction at room temperature and undergoes a double
transition to a novel Peierls insulator with decreasing temperature.
\cite{K10068,Y1198}
($\mu$-bpym)[Pt(en)$X$]$_2X$(ClO$_4$)$_3\cdot$H$_2$O
($X=\,$Cl, Br;
 $\mu$-bpym$\,=2,2'$-bipyrimidine$\,=\,$C$_8$H$_6$N$_4$) \cite{K7372}
and
(bpy)[Pt(dien)Br]$_2$Br$_4\cdot 2$H$_2$O
(dien$\,=\,$diethylentriamine$\,=\,$C$_4$H$_{13}$N$_3$) \cite{K12066}
were synthesized in an attempt to bring a couple of $M\!X$ chains into
interaction.
They are made in similar ladder structures but in distinct ground states
of mixed valence, \cite{F044717,I063708} which are optically
distinguishable. \cite{Y235116,Y367}
\begin{figure}
\centering
\includegraphics[width=85mm]{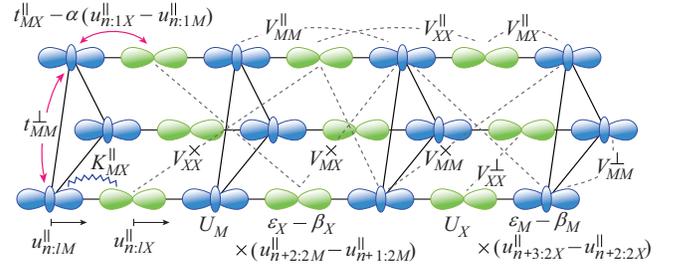}
\vspace*{-2mm}
\caption{(Color online)
         Modeling of an $M\!X$ triangular prism, where heavily and
         lightly shaded clouds denote $M\,d_{z^2}$ and $X\,p_z$
         orbitals, the electron numbers on which are calculated through
         $n_{n:lMs}\equiv a_{n:lMs}^\dagger a_{n:lMs}$ and
         $n_{n:lXs}\equiv a_{n:lXs}^\dagger a_{n:lXs}$, respectively.
         The on-site energies (electron affinities) of isolated atoms
         are given by $\varepsilon_M$ and $\varepsilon_X$, while the
         electron hops between these levels are expressed by
         $t_{M\!X}^{\parallel}$ and $t_{M\!M}^{\perp}$.
         The on-site Coulomb interactions are labeled as $U_A$ ($A=M,X$),
         whereas the interchain and intrachain different-site Coulomb
         interactions as
         $V_{A\!A'}^{\perp}$,
         $V_{A\!A'}^{\times}$ and
         $V_{A\!A'}^{\parallel}$ ($A,A'=M,X$).
         The leg-direction displacements of metal and halogen ions,
         $u_{n:lM}^{\parallel}$ and $u_{n:lX}^{\parallel}$, interact with
         electrons through the intersite ($\alpha$) and intrasite
         ($\beta_M,\beta_X$) coupling constants at the cost of elastic
         energy due to the spring constant $K_{M\!X}^{\parallel}$.}
\label{F:H}
\end{figure}

   Thus and thus, the $M\!X$ class of materials have been fascinating
numerous chemists and physicists for more than half a century.
Much effort is still devoted to elaborating new varieties and the novel
porous nanotube [Pt(en)(bpy)I]$_4$(NO$_3$)$_8$ has just been fabricated.
Diffuse X-ray scattering measurements \cite{Ounpub} on it suggest two or
more cell-doubled mixed-valent phases competing with each other in the
ground state.
Such phases are indeed recognizable as broken-symmetry solutions of
a four-legged Peierls-Hubbard Hamiltonian of tetragonal symmetry.
\cite{O17006}
Then we may have an idea of bringing coupled $M\!X$ chains into geometric
frustration.
A platinum-halide triangular-prism compound must be a unique
charge-frustrated nanotubular system and may exhibit novel valence
arrangements which have never been observed in any other material.
Thus motivated, we apply a group-theoretical bifurcation theory
\cite{N413,H391} to a three-legged Peierls-Hubbard Hamiltonian of
hexagonal symmetry and reveal its exotic ground states.
They are numerically calculated and compared with each other
in an attempt to guide future experiments.

\section{Model Hamiltonian and Its Symmetry Properties}

   Metal-halide triangular prisms are describable in terms of a two-band
extended Peierls-Hubbard Hamiltonian of hexagonal symmetry,
\begin{eqnarray}
   &&\!\!\!\!\!\!\!\!
   {\cal H}=
   \sum_{l,n,s}
   \Bigl\{
    \bigl[t_{M\!X}^{\parallel}
         -\alpha(u_{n+1:lM}^{\parallel}-u_{n:lX}^{\parallel})\bigr]
    a^{\dagger}_{n+1:lMs}a_{n:lXs}
   \nonumber \\
   &&\!\!\!\!\!\!\!\!\qquad\quad
   -\bigl[t_{M\!X}^{\parallel}
         -\alpha(u_{n:lX}^{\parallel}-u_{n:lM}^{\parallel})\bigr]
    a^{\dagger}_{n:lXs}a_{n:lMs}
   \nonumber \\
   &&\!\!\!\!\!\!\!\!\qquad\quad
   -t_{M\!M}^{\perp}
    a^{\dagger}_{n:l+1Ms}a_{n:lMs}
   +{\rm H.c.}
   \Bigr\}
   \nonumber \\
   &&\!\!\!\!\!\!\!\!\quad
  +\sum_{l,n,s}
   \Bigl\{
    \bigl[\varepsilon_{M}
         -\beta_{M}(u_{n:lX}^{\parallel}-u_{n-1:lX}^\parallel)\bigr]
    n_{n:lMs}
   \nonumber \\
   &&\!\!\!\!\!\!\!\!\qquad\quad
   +\bigl[\varepsilon_{X}
         -\beta_{X}(u_{n+1:lM}^{\parallel}-u_{n:lM}^{\parallel})\bigr]
    n_{n:lXs}
   \Bigr\}
   \nonumber \\
   &&\!\!\!\!\!\!\!\!\quad
  +\sum_{l,n}\frac{K_{M\!X}^{\parallel}}{2}
    \bigl[(u_{n:lX}^{\parallel}-u_{n:lM}^{\parallel})^{2}
         +(u_{n+1:lM}^{\parallel}-u_{n:lX}^{\parallel})^{2}\bigr]
   \nonumber \\
   &&\!\!\!\!\!\!\!\!\quad
  +\sum_{A=M,X}\sum_{l,n,s,s'}
   \Bigl\{
    \frac{U_{A}}{4}n_{n:lAs}n_{n:lA-s}
   +V_{A\!A}^{\parallel}n_{n:lAs}n_{n+1:lAs'}
   \nonumber \\
   &&\!\!\!\!\!\!\!\!\qquad\quad
   +V_{A\!A}^{\times}
    (n_{n:lAs}n_{n+1:l+1As'}+n_{n:l+1As}n_{n+1:lAs'})
   \nonumber \\
   &&\!\!\!\!\!\!\!\!\qquad\quad
   +V_{A\!A}^{\perp}n_{n:lAs}n_{n:l+1As'}
   \Bigr\}
   \nonumber \\
   &&\!\!\!\!\!\!\!\!\quad
  +\sum_{l,n,s,s'}
   \Bigl\{
    V_{M\!X}^{\parallel}
    (n_{n:lMs}n_{n:lXs'}+n_{n:lXs}n_{n+1:lMs'})
   \nonumber \\
   &&\!\!\!\!\!\!\!\!\qquad\quad
   +V_{M\!X}^{\times}
    (n_{n:lMs}n_{n:l+1Xs'}+n_{n:l+1Ms}n_{n:lXs'}
   \nonumber \\
   &&\!\!\!\!\!\!\!\!\qquad\quad
    +n_{n:lXs}n_{n+1:l+1Ms'}+n_{n:l+1Xs}n_{n+1:lMs'})
   \Bigr\},
   \label{E:H}
\end{eqnarray}
where $M\!X$ chain legs and $M_3X_3$ prism units are numbered by $l=1,2,3$
and $n=1,2,\cdots,N$, respectively, while electron spins are indicated by
$s,s'=\uparrow,\downarrow$.
The modeling is visualized and explained in more detail in Fig. \ref{F:H}. 

   Unless the gauge symmetry is broken, the symmetry group of any lattice
electron system can be written as
$\mathbf{G}=\mathbf{P}\times\mathbf{S}\times\mathbf{T}$,
where $\mathbf{P}$, $\mathbf{S}$ and $\mathbf{T}$ are the groups of
space, spin rotation, and time reversal, respectively.
The space group is further decomposed into the translation and point
groups as $\mathbf{L}\land\mathbf{D}$.
For $M\,d_{z^2}$-$X\,p_z$ triangular prisms, $\mathbf{L}$ and $\mathbf{D}$
read as $\{E,l\}\equiv\mathbf{L}_1$ and $\mathbf{D}_{3h}$,
respectively, where $l$ is the unit-cell translation in the $z$ direction.
Defining the Fourier transformation as
$a_{k:lAs}
 =N^{-1/2}\sum_n e^{-ik(n+\delta_{AX}/2)}a_{n:lAs}$ and
$u_{k:lA}^{\parallel}
 =N^{-1/2}\sum_n e^{-ik(n+\delta_{AX}/2)}u_{n:lA}^{\parallel}$ with the
lattice constant along the tube axis set equal to unity and composing
Hermitian bases of the gauge-invariant operators
$\{a_{k:lAs}^{\dagger}a_{k':l'A's'}\}$,
we find out irreducible representations of $\mathbf{G}$ over the
real number field, which are referred to as $\check{G}$.
Actions of $l\in\mathbf{L}_1$ and $t\in\mathbf{T}$ on the electron
operators are defined as
$l\cdot a_{k:lAs}^\dagger=e^{-ikl}a_{k:lAs}^\dagger$ and
$t\cdot a_{k:lAs}^\dagger
 =(-1)^{\delta_{s\uparrow}}a_{-k:lA-s}^\dagger$.
Those of $p\in\mathbf{D}_{3h}$ are calculated as
$p\cdot a_{k:lMs}^\dagger=[A_{1}'(p)]_{11}a_{pk:lMs}^\dagger$ and
$p\cdot a_{k:lXs}^\dagger=[A_{2}''(p)]_{11}a_{pk:lXs}^\dagger$, where
$[\check{D}(p)]_{ij}$ is the $(i,j)$-element of the $\check{D}$
representation matrix for $p$.
Those of
$u(\mbox{\boldmath$e$},\theta)
 =\sigma^0\cos(\theta/2)
 -i(\mbox{\boldmath$\sigma$}\cdot\mbox{\boldmath$e$})
  \sin(\theta/2)
 \in\mathbf{S}$
are represented as
$u(\mbox{\boldmath$e$},\theta)\cdot a_{k:lAs}^\dagger
 =\sum_{s'}[u(\mbox{\boldmath$e$},\theta)]_{s's}a_{k:lAs'}^\dagger$,
where $\sigma^0$ and
$\mbox{\boldmath$\sigma$}=(\sigma^x,\sigma^y,\sigma^z)$
are the $2\times 2$ unit matrix and a vector composed of the Pauli
matrices, respectively.
Any representation $\check{G}$ is expressed as
$\check{G}=\check{P}\otimes\check{S}\otimes\check{T}$.
Once a wave vector $Q$ is fixed, the relevant little group
$\mathbf{D}(Q)$ is given.
$\check{P}$ is therefore labeled as $Q\check{D}(Q)$.
The relevant representations of $\mathbf{S}$ are given by
$\check{S}^0(u(\mbox{\boldmath$e$},\theta))
 =1$ (singlet) and
$\check{S}^1(u(\mbox{\boldmath$e$},\theta))
 =O(u(\mbox{\boldmath$e$},\theta))$ (triplet),
where $O(u(\mbox{\boldmath$e$},\theta))$ is the $3\times 3$
orthogonal matrix satisfying
$
   u(\mbox{\boldmath$e$},\theta)
   \mbox{\boldmath$\sigma$}^\lambda
   u^\dagger(\mbox{\boldmath$e$},\theta)
      =\sum_{\mu=x,y,z}
       [O(u(\mbox{\boldmath$e$},\theta))]_{\lambda\mu}
       \mbox{\boldmath$\sigma$}^\mu \ \
       (\lambda=x,\,y,\,z)
$.
Those of $\mathbf{T}$ are given by
$\check{T}^0(t)=1$ (symmetric) and $\check{T}^1(t)=-1$ (antisymmetric).
Considering that platinum $5d$ electrons are moderately correlated,
magnetically broken-symmetry solutions are less interesting.
Nontrivial current-wave phases, whether of charge \cite{Y329} or of spin,
\cite{Y335} are of little occurrence unless electrons are well itinerant
in both leg and rung directions.
Therefore, we discuss density-wave solutions of the
$\check{P}\otimes\check{S}^0\otimes\check{T}^0$ type.
Since the relevant $d_{z^2}$ and $p_z$ orbitals of constituent $M\!X$
chains are formally half and fully filled, respectively, we take much
interest in cell-doubled mixed-valent states among others.
Increasing temperature and/or electrochemical doping may suppress lattice
dimerization and induce valence delocalization.
Thus we consider the cases of $Q=0$ and $Q=\pi$, which are described by
the translation groups
$\mathbf{L}_1\equiv\{E,l\}$ and $\mathbf{L}_2\equiv\{E,2l\}$,
respectively, and are hereafter referred to as $\Gamma$ and ${\rm X}$,
respectively.
Both $\mathbf{D}(\Gamma)$ and $\mathbf{D}({\rm X})$ read as
$\mathbf{D}_{3h}$.
\begin{table*}
\caption{Symmetry properties of irreducible representations,
$\Gamma\check{D}(\Gamma)\otimes\check{S}^0\otimes\check{T}^0$ and
${\rm X}\check{D}({\rm X})\otimes\check{S}^0\otimes\check{T}^0$,
available on the condition of axial isotropy subgroup.}
\begin{tabular}{ccccc}
\hline \hline
  $\begin{matrix}
  \text{Irreducible} \\[-1.0mm]
  \text{representation}
  \end{matrix}$
& $\begin{matrix}
  \text{Axial isotropy}\\[-1.0mm]
  \text{subgroup}
  \end{matrix}$
& $\begin{matrix}
  \text{Fixed-point}\\[-1.0mm]
  \text{subspace}
  \end{matrix}$
& $\begin{matrix}
  \text{Broken-symmetry}\\[-1.0mm]
  \text{Hamiltonian}
  \end{matrix}$
& $\begin{matrix}
  \text{Physical}\\[-1.0mm]
  \text{character}
  \end{matrix}$ \\[2.7mm]
\hline
  ${\Gamma}A'_{1}\otimes\check{S}^{0}\otimes\check{T}^{0}$
& $\mathbf{D}_{3h}\mathbf{L}_{1}\mathbf{ST}$
& $h_{{\Gamma}A_{1}'[1,1]}^{00}$
& $h_{{\Gamma}A_{1}'[1,1]}^{00}$
& PM
\\
  ${\Gamma}A'_{2}\otimes\check{S}^{0}\otimes\check{T}^{0}$
& $\mathbf{C}_{3h}\mathbf{L}_{1}\mathbf{ST}$
& $h_{{\Gamma}A_{2}'[1,1]}^{00}$
& $h_{{\Gamma}A_{2}'[1,1]}^{00}
  +h_{{\Gamma}A_{1}'[1,1]}^{00}$
& $\vee$-$M\!X$-BOW
\\
  ${\Gamma}A''_{1}\otimes\check{S}^{0}\otimes\check{T}^{0}$
& $\mathbf{D}_{3}\mathbf{L}_{1}\mathbf{ST}$
& $h_{{\Gamma}A_{1}''[1,1]}^{00}$
& $h_{{\Gamma}A_{1}''[1,1]}^{00}
  +h_{{\Gamma}A_{1}'[1,1]}^{00}$
& $\setminus \! \setminus$-$M\!X$-BOW
\\
  ${\Gamma}A''_{2}\otimes\check{S}^{0}\otimes\check{T}^{0}$
& $\mathbf{C}_{3v}\mathbf{L}_{1}\mathbf{ST}$
& $h_{{\Gamma}A_{2}''[1,1]}^{00}$
& $h_{{\Gamma}A_{2}''[1,1]}^{00}
  +h_{{\Gamma}A_{1}'[1,1]}^{00}$
& $(---)$-$M\!X$-BOW
\\
  ${\Gamma}E'(1)\otimes\check{S}^{0}\otimes\check{T}^{0}$
& $\mathbf{C}_{2v}\mathbf{L}_{1}\mathbf{ST}$
& $h_{{\Gamma}E'[1,1]}^{00}$
& $h_{{\Gamma}E'[1,1]}^{00}
  +h_{{\Gamma}A_{1}'[1,1]}^{00}$
& $\bigtriangleup$-$M\!M$-BOW
\\
  ${\Gamma}E''(1)\otimes\check{S}^{0}\otimes\check{T}^{0}$
& $(1+\sigma_{v1})\mathbf{L}_{1}\mathbf{ST}$
& $h_{{\Gamma}E''[1,1]}^{00}$
& $h_{{\Gamma}E''[1,1]}^{00}
  +h_{{\Gamma}A_{2}''[1,1]}^{00}
  +h_{{\Gamma}E'[1,1]}^{00}
  +h_{{\Gamma}A_{1}'[1,1]}^{00}$
& $(\pm{\mbox{\large \boldmath$-$}}{\mbox{\large \boldmath$-$}})$-$M\!X$-BOW
\\
  ${\Gamma}E''(2)\otimes\check{S}^{0}\otimes\check{T}^{0}$
& $(1+C'_{21})\mathbf{L}_{1}\mathbf{ST}$
& $h_{{\Gamma}E''[2,2]}^{00}$
& $h_{{\Gamma}E''[2,2]}^{00}
  +h_{{\Gamma}A_{1}''[1,1]}^{00}
  +h_{{\Gamma}E'[1,1]}^{00}
  +h_{{\Gamma}A_{1}'[1,1]}^{00}$
& $(0-+)$-$M\!X$-BOW
\\
  X$A'_{1}\otimes\check{S}^{0}\otimes\check{T}^{0}$
& $\mathbf{D}_{3h}\mathbf{L}_{2}\mathbf{ST}$
& $h_{{\rm X}A_{1}'[1,1]}^{00}$
& $h_{{\rm X}A_{1}'[1,1]}^{00}
  +h_{{\Gamma}A_{1}'[1,1]}^{00}$
& $(+++)$-$M$-CDW
\\
  X$A'_{2}\otimes\check{S}^{0}\otimes\check{T}^{0}$
& $(1+C'_{21}l)\mathbf{C}_{3h}\mathbf{L}_{2}\mathbf{ST}$
& $h_{{\rm X}A_{2}'[1,1]}^{00}$
& $h_{{\rm X}A_{2}'[1,1]}^{00}
  +h_{{\Gamma}A_{1}'[1,1]}^{00}$
& $\times$-$M\!M$-BOW
\\
  X$A_{1}''\otimes\check{S}^{0}\otimes\check{T}^{0}$
& $(1+\sigma_{h}l)\mathbf{D}_{3}\mathbf{L}_{2}\mathbf{ST}$
& $h_{{\rm X}A_{1}''[1,1]}^{00}$
& $h_{{\rm X}A_{1}''[1,1]}^{00}
  +h_{{\Gamma}A_{1}'[1,1]}^{00}$
& $\times$-$X\!X$-BOW
\\
  X$A_{2}''\otimes\check{S}^{0}\otimes\check{T}^{0}$
& $(1+C_{21}'l)\mathbf{C}_{3v}\mathbf{L}_{2}\mathbf{ST}$
& $h_{{\rm X} A_{2}''[1,1]}^{00}$
& $h_{{\rm X}A_{2}''[1,1]}^{00}
  +h_{{\Gamma}A_{1}'[1,1]}^{00}$
& $(+++)$-$X$-CDW
\\
  X$E'(1)\otimes\check{S}^{0}\otimes\check{T}^{0}$
& $\mathbf{C}_{2v}\mathbf{L}_{2}\mathbf{ST}$
& $h_{{\rm X}E'[1,1]}^{00}$
& $h_{{\rm X}E'[1,1]}^{00}
  +h_{{\rm X}A_{1}'[1,1]}^{00}
  +h_{{\Gamma}E'[1,1]}^{00}
  +h_{{\Gamma}A_{1}'[1,1]}^{00}$
& $({\mbox{\large \boldmath$\pm$}}++)$-$M$-CDW
\\
  X$E'(2)\otimes\check{S}^{0}\otimes\check{T}^{0}$
& $(1+C'_{21}l)\mathbf{C}_{1h}\mathbf{L}_{2}\mathbf{ST}$
& $h_{{\rm X} E'[2,2]}^{00}$
& $h_{{\rm X}E'[2,2]}^{00}
  +h_{{\rm X}A_{2}'[1,1]}^{00}
  +h_{{\Gamma}E'[1,1]}^{00}
  +h_{{\Gamma}A_{1}'[1,1]}^{00}$
& $(0+-)$-$M$-CDW
\\
  X$E''(1)\otimes\check{S}^{0}\otimes\check{T}^{0}$
& $(1+C_{21}'l)(1+\sigma_{v1})\mathbf{L}_{2}\mathbf{ST}$
& $h_{{\rm X} E''[1,1]}^{00}$
& $h_{{\rm X}E''[1,1]}^{00}
  +h_{{\rm X}A_{2}''[1,1]}^{00}
  +h_{{\Gamma}E'[1,1]}^{00}
  +h_{{\Gamma}A_{1}'[1,1]}^{00}$
& $({\mbox{\large \boldmath$\pm$}}++)$-$X$-CDW
\\
  X$E''(2)\otimes\check{S}^{0}\otimes\check{T}^{0}$
& $(1+C_{21}')(1+\sigma_{v1}l)\mathbf{L}_{2}\mathbf{ST}$
& $h_{{\rm X} E''[2,2]}^{00}$
& $h_{{\rm X}E''[2,2]}^{00}
  +h_{{\rm X}A_{1}''[1,1]}^{00}
  +h_{{\Gamma}E'[1,1]}^{00}
  +h_{{\Gamma}A_{1}'[1,1]}^{00}$
& $(0+-)$-$X$-CDW
\\[1.0mm]
\hline \hline
\end{tabular}
\label{T:Rrep}
\end{table*}

\begin{figure*}
\centering
\includegraphics[width=160mm]{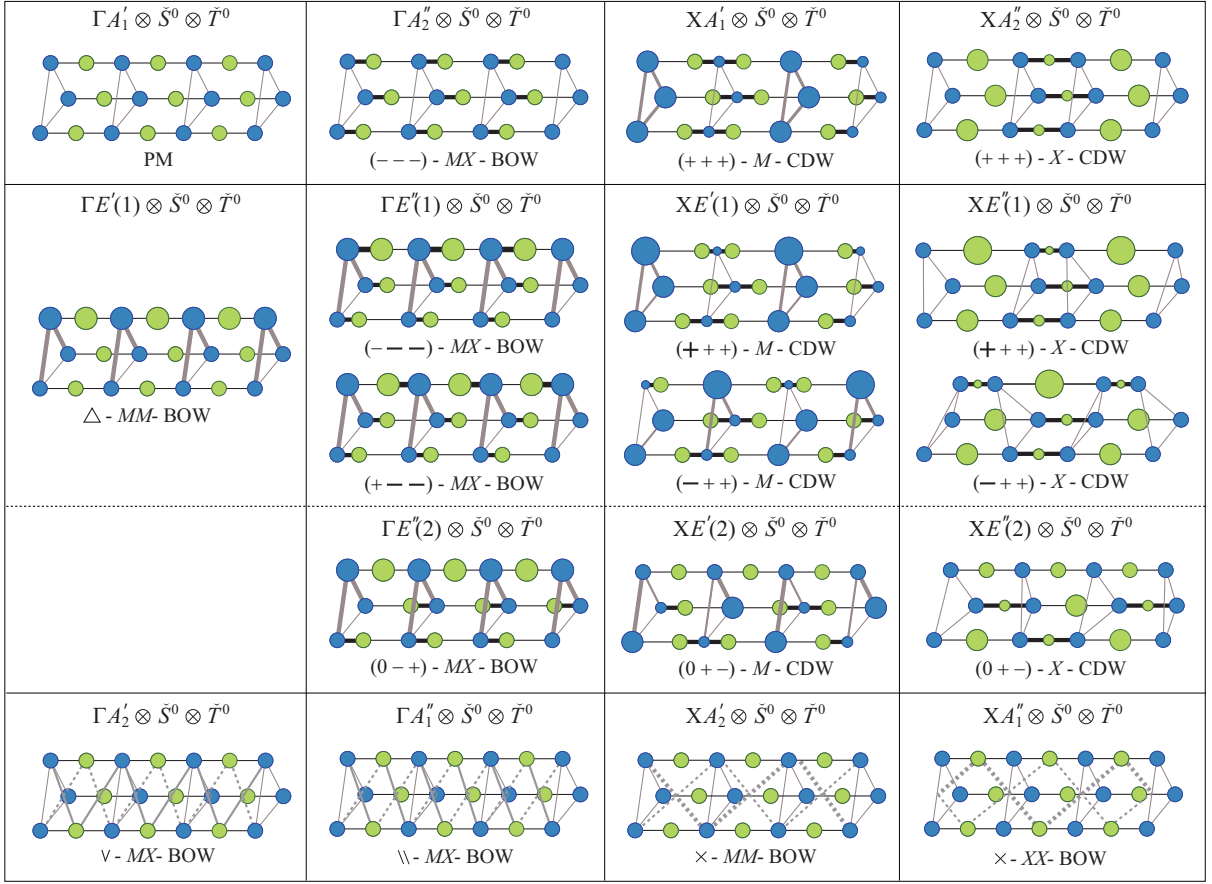}
\vspace*{-2mm}
\caption{(Color online)
         Charge-density-wave (CDW) and bond-order-wave (BOW) states
         obtained from the irreducible representations listed in
         Table \ref{T:Rrep}, where varied circles and segments represent
         oscillating electron densities and bond orders, respectively,
         while irregularly arranged circles denote lattice distortion.
         Various $M$ ($X$)-CDW states are referred to as
         $(\sigma_1,\sigma_2,\sigma_3)$, which signify the electron
         densities on adjacent metal (halogen) sites forming a triangular
         section of the prism and are measured in comparison with the
         average occupancy of each chain $l$.
         Some of BOW states are similarly nicknamed, where the signatures
         denote the length changes from the equidistant spacing of
         adjacent $M\!X$ bonds forming a prism fragment.
         Coexistent normal ($\pm$) and larger accentuated
         ($\mbox{\large \boldmath$\pm$}$) signatures represent
         quantitatively different amplitudes.}
\label{F:DW}
\end{figure*}

   We take the Hartree-Fock scheme of rewriting the Hamiltonian
(\ref{E:H}) into
\begin{eqnarray}
   &&\!\!\!\!
   {\cal H}_{\rm HF}
   =\sum_{l,l'}\sum_{A,A'}
    \sum_{Q=\mathit{\Gamma},{\rm X}}\sum_{k,s,s'}
    \sum_{\lambda=0,x,y,z}
    x_{lAl'A'}^\lambda(Q;k)
   \nonumber\\
   &&\qquad\times
    a_{k+Q:lAs}^\dagger a_{k:l'A's'}\sigma_{ss'}^\lambda
   \equiv\sum_{Q}\sum_\lambda h_Q^\lambda,
   \label{E:HHF}
\end{eqnarray}
where the self-consistent fields $x_{lAl'A'}^\lambda(Q;k)$, as well as
the lattice distortions $u_{Q:lA}^\parallel$, can be described in terms
of the density matrices
$\rho_{l'A'lA}^\lambda(Q;k)
 =\sum_{s,s'}\langle a_{k+Q:lAs}^\dagger a_{k:l'A's'}\rangle_T
  \sigma_{ss'}^\lambda/2$ with $\langle\cdots\rangle_T$ denoting the
canonical average and are adiabatically determined at each temperature $T$
so as to minimize the free energy.
Employing the projection operators
\begin{equation}
   P_{\check{D}[i,j]}^\tau
  =\frac{d_{\check{D}}}{2g}
    \sum_{t\in\mathbf{T}}\check{T}^\tau(t)
    \sum_{p\in\mathbf{D}_{3h}}[\check{D}(p)]_{ij}^*tp,
\end{equation}
where
$g\,(=12)$ is the order of $\mathbf{D}_{3h}$ and
$d_{\check{D}}\,(\leq 2)$ is the dimension of its arbitrary irreducible
representation $\check{D}$, we further decompose the Hamiltonian
(\ref{E:HHF}) into its fixed-point subspaces as
\begin{equation}
   \!\!\!\!
   {\cal H}_{\rm HF}
   =\sum_{Q=\mathit{\Gamma},X}
    \sum_{\check{D}(Q)}
    \sum_{\lambda=0,x,y,z}
    \sum_{\tau=0,1}
    h_{Q\check{D}(Q)}^{\lambda\tau},
\end{equation}
where
$h_{Q\check{D}[i,j]}^{\lambda\tau}
=P_{\check{D}[i,j]}^\tau\cdot h_Q^\lambda$.
We list in Table \ref{T:Rrep} the thus-obtained irreducible
representations together with their broken-symmetry Hamiltonians.
The two-dimensional representations $E'$ and $E''$ are generally available
in a variety of isotropy subgroups.
Here we consider only the axial ones discarding those of two-dimensional
fixed point subspace.
Every irreducible representation is guaranteed to yield a stable solution
only when its isotropy subgroup possesses a one-dimensional fixed point
subspace. \cite{G83}
Considering that the density matrices are of the same symmetry as their
host Hamiltonian, we learn the oscillating pattern of electron densities,
$\sum_s
 \langle a_{n:lAs}^\dagger a_{n:lAs} \rangle_T
 \equiv d_{n:lA}$,
and bond orders,
$\mbox{Re}\sum_s
 \langle a_{n:lAs}^\dagger a_{n':l'A's} \rangle_T
 \equiv p_{n:lA;n':l'A'}$.
In Fig. \ref{F:DW} we draw and name the consequent charge-density-wave
(CDW) solutions, including bond-order-wave (BOW) ones which are
recognizable as bond-centered CDW states.

   Two kinds of axial isotropy subgroups are available from all the
two-dimensional representations but $\Gamma E'$.
A single state is derived from each of the second kind, whereas two
quantitatively different states can be born to each of the first kind,
which are distinguishably nicknamed.
These symmetrically (qualitatively) degenerate but practically
(quantitatively) distinct varieties read as heterogeneous CDW states,
where each chain is no longer $3/4$-filled, that is, one or two chains are
charge-rich, while the rest are charge-poor.
Such an interchain charge polarization is common to all the two-dimensional
representations, including
${\Gamma}E'(1)\otimes\check{S}^{0}\otimes\check{T}^{0}$.
The thus-broken electron-hole symmetry yields an exotic phase diagram.
It is the geometric frustration rather than any orbital hybridization that
breaks down the electron-hole symmetry in triangular-prism $M\!X$
complexes.
It is not the case with even-legged ladders \cite{Y235116} and prisms,
\cite{O17006} where the electron-hole symmetry is kept within any
single-band description and the $d$-electron phase diagram as a function
of doping is symmetric with respect to the half occupancy.

   All the $\Gamma\check{D}(\Gamma)\otimes\check{S}^0\otimes\check{T}^0$
solutions but the paramagnetic metal (PM) of the full symmetry
$\mathbf{D}_{3h}\mathbf{L}_{1}\mathbf{ST}$ are characterized as BOW
states.
There is no lattice distortion in some of them under the present
modeling, but even $\bigtriangleup$-$M\!M$-BOW may be accompanied by cell
deformation on the assumption that the interchain electron transfer
$t_{M\!M}^{\perp}$ can be coupled to phonons.
Every BOW may be stabilized by electrons directly hopping on the
oscillating bonds and their interactions with phonons, but any is of
little occurrence within realistic modeling and parametrization.
BOW states are more likely to appear in decoupled chains of the $M\!X$
\cite{B339,Y422} and $M\!M\!X$ \cite{Y140102,Y1198} types.

   The ${\rm X}\check{D}({\rm X})\otimes\check{S}^0\otimes\check{T}^0$
solutions are classified into three groups:
CDWs on the metal sublattice with the halogen sublattice dimerized,
those on the halogen sublattice with the metal sublattice dimerized and
BOWs without any charge oscillation.
We find twice three irreducible representations assuming a CDW character.
There is a one-to-one correspondence between $M$- and $X$-CDWs.
$X$-CDW states consist of mixed-valent halogen ions.
$p$ electrons may be activated with increasing $\varepsilon_X$ and $U_X$,
but platinum-fluorides are hard to fabricate.
Although all CDW states gain a condensation energy due to the Peierls
distortion, they are not necessarily gapped.
Those of the $(0+-)$ type are {\it partially metallic}, where only
two chains are valence-trapped and the rest is valence-delocalized.
Such states as cell-doubled but gapless at the Fermi level are never
available from $M\!X$ ladders \cite{F044717} but generally possible in
$M\!X$ tubes, whose little groups $\mathbf{D}({\rm X})$ necessarily have a
two-dimensional irreducible representation of axial isotropy subgroup.
All other CDW states are fully gapped at the boundaries of the reduced
Brillouin zone.
\begin{figure*}
\centering
\includegraphics[width=160mm]{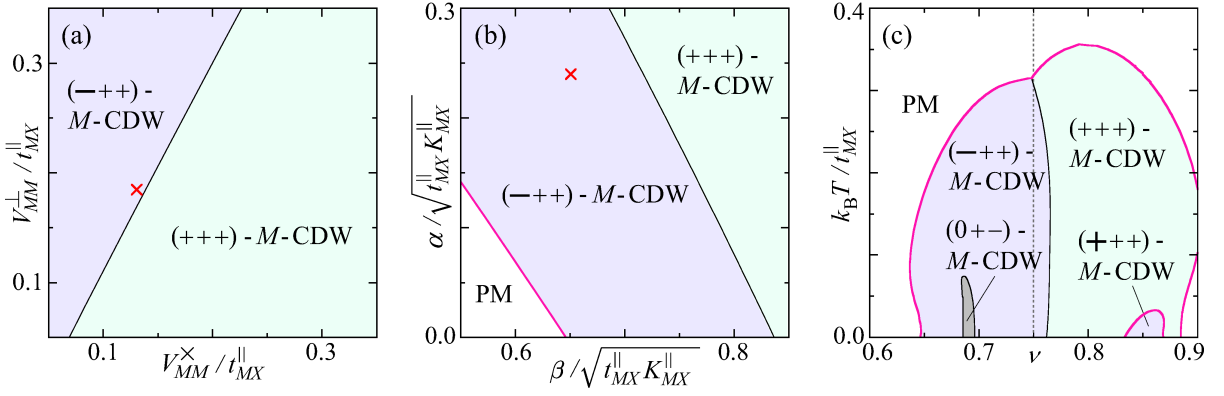}
\vspace*{-2mm}
\caption{(Color online)
         (a) A ground-state phase diagram with competing electron-electron
             interactions, $V_{M\!M}^{\times}$ and $V_{M\!M}^{\perp}$;
         (b) A ground-state phase diagram with varying electron-lattice
             couplings, $\alpha$ and $\beta\equiv\beta_{M}=\beta_{X}$;
         (c) A thermal phase diagram as a function of the electron
             occupancy $\nu$, where the dotted line of $\nu=3/4$ is a
             guide for eyes, separating the hole- and electron-doped
             regions.
         In all the diagrams the crosses indicate the standard
         parametrization stated in the text and we move away from this
         point tuning only the parameters in issue.
         Phase boundaries drawn in black and colored red (or toned down
         in monochromes) denote transitions of the first and second
         order, respectively.}
\label{F:PhD}
\end{figure*}

\section{Quantum and Thermal Phase Diagrams}

   Now we numerically draw various phase diagrams.
We model a trial $M\!X$ prism on the platinum-halide ladder compound
($\mu$-bpym)[Pt(en)$X$]$_2X$(ClO$_4$)$_3\cdot$H$_2$O. \cite{K7372}
Assuming that platinum ions are equally spaced in the leg and rung
directions, $r_{M\!M}^\parallel=2r_{M\!X}^\parallel=r_{M\!M}^\perp$,
we set the transfer integrals for $t_{M\!M}^\perp=0.4t_{M\!X}^\parallel$.
The on-site electronic parameters are taken as
$U_M=0.80t_{M\!X}^\parallel$, $U_X=0.66t_{M\!X}^\parallel$ and
$\varepsilon_M-\varepsilon_X=2.0t_{M\!X}^\parallel$, while the elastic
constants are adjusted to
$\alpha=0.24\sqrt{t_{M\!X}^\parallel K_{M\!X}^\parallel}$
and $\beta_M=\beta_X=0.65\sqrt{t_{M\!X}^\parallel K_{M\!X}^\parallel}$.
Coulomb interactions between different sites are designed as
$V_{M\!X}^\parallel=(U_M+U_X)/4$,
$V_{A\!A}^\parallel
=V_{M\!X}^\parallel r_{M\!X}^\parallel/r_{A\!A}^\parallel$,
$V_{A\!A}^\perp
=V_{M\!X}^\parallel r_{M\!X}^\parallel/r_{A\!A}^\perp$ and
$V_{A\!A'}^\times
=V_{M\!X}^\parallel r_{M\!X}^\parallel
/\sqrt{(r_{M\!M}^\perp)^2+(r_{A\!A'}^\parallel)^2}$.
Such a parametrization is so realistic as to successfully interpret
optical observations of existent platinum-halide two-leg ladders
\cite{Y235116} and quadratic prisms. \cite{O17006}
We employ these parameters unless otherwise noted, which are referred to
as the standard parametrization and are indicated by $\times$ in
ground-state phase diagrams.
\begin{figure}
\centering
\includegraphics[width=85mm]{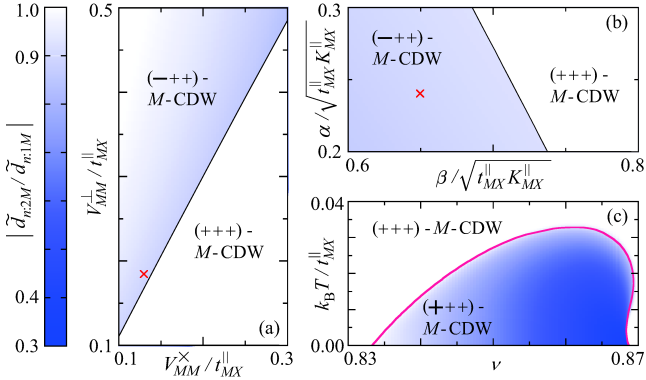}
\vspace*{-2mm}
\caption{(Color online)
         Contour plots of the ratio between the CDW amplitudes in the first
         and second chains on Figs. \ref{F:PhD}(a), \ref{F:PhD}(b) and
         \ref{F:PhD}(c) magnified.}
\label{F:contour}
\end{figure}

   Figure \ref{F:PhD} shows quantum and thermal phase competitions
under varying electron-electron interactions, electron-lattice couplings
and electron occupancy.
Ground-state phase diagrams are calculated at a sufficiently low
temperature, $k_{\rm B}T/t_{M\!X}^\parallel=0.001$.
In $({\mbox{\large \boldmath$-$}}++)$-$M$-CDW and
$({\mbox{\large \boldmath$+$}}++)$-$M$-CDW, three chains may unequally
be distorted, that is, the CDW amplitude of one chain may be larger
than those of the rest two.
In order to detect such a heterogeneous CDW, we decompose the electron
densities $d_{n:lM}$ into net (${\bar d}_{n:lM}$) and alternating
(${\widetilde d}_{n:lM}$) components as \cite{Y165113}
\begin{eqnarray}
   &&\!\!\!\!
   d_{n:lM}={\bar d}_{n:lM}+{\widetilde d}_{n:lM};
   \nonumber \\
   &&\!\!\!\!
   {\widetilde d}_{n:lM}
  =\frac{1}{4}\bigl(2d_{n:lM}-d_{n-1:lM}-d_{n+1:lM}\bigr),
\end{eqnarray}
and calculate the ratios $|{\widetilde d}_{n:2M}/{\widetilde d}_{n:1M}|$
(Fig. \ref{F:contour}).
These ratios are independent of the unit index $n$ in any ground state and
correspond to unity when three chains are proportionate in charge.
$({\mbox{\large \boldmath$+$}}++)$-$M$-CDW, bifurcating from
$(+++)$-$M$-CDW, is possibly of remarkable interchain charge polarization,
where the interchain heterogeneity barometer
$|{\widetilde d}_{n:2M}/{\widetilde d}_{n:1M}|$ goes almost down to $0.3$,
while $({\mbox{\large \boldmath$-$}}++)$-$M$-CDW, discontinuously
replacing $(+++)$-$M$-CDW, is generally of moderate interchain charge
polarization, where $|{\widetilde d}_{n:2M}/{\widetilde d}_{n:1M}|$ is not
so far from unity.
Then we may have an idea of describing the
$({\mbox{\large \boldmath$-$}}++)$-to-$(+++)$-$M$-CDW transition in terms
of weakly coupled CDW chains of Pt$^{2+}$ and Pt$^{4+}$.

   Figure \ref{F:PhD}(a) demonstrates that $V_{M\!M}^{\times}$ and
$V_{M\!M}^{\perp}$ are the driving interactions for $(+++)$-$M$-CDW and
$({\mbox{\large \boldmath$-$}}++)$-$M$-CDW, respectively.
Under strong valence localization, their $d$-electron energies may be
expressed as
\begin{eqnarray}
   &&\!\!\!\!
   \frac{E_{(+++)\mbox{-}M\mbox{-}{\rm CDW}}}{N}
  =3\varepsilon_M-\frac{3\beta_M^2}{K_{M\!X}^\parallel}+\frac{3U_M}{2}
   \nonumber \\
   &&\!\!\!\!\qquad
  +6V_{M\!M}^\perp,
   \label{E:E(+++)}
   \\
   &&\!\!\!\!
   \frac{E_{(-++)\mbox{-}M\mbox{-}{\rm CDW}}}{N}
  =3\varepsilon_M-\frac{3\beta_M^2}{K_{M\!X}^\parallel}+\frac{3U_M}{2}
   \nonumber \\
   &&\!\!\!\!\qquad
  +2V_{M\!M}^\perp+8V_{M\!M}^\times,
   \label{E:E(-++)}
\end{eqnarray}
which are balanced at $V_{M\!M}^\perp=2V_{M\!M}^\times$ and are consistent
with numerical findings in Fig. \ref{F:PhD}(a) to a certain extent.
We can refine such an analytic consideration taking account of transfer
effects.
Decoupled chains, whether in $(+++)$-$M$-CDW or in $(-++)$-$M$-CDW,
gain the same hopping energy on the zeroth-order phase boundary
$V_{M\!M}^\perp=2V_{M\!M}^\times$.
The interchain contact $t_{M\!M}^\perp$ switched on contributes further
stabilization energy to the $(-++)$ arrangement but is of no benefit to
the $(+++)$ one.
We obtain a correcter estimate of the phase boundary as
\begin{eqnarray}
   &&\!\!\!\!\!\!\!\!\!\!\!\!\!\!\!\!\!\!
   V_{M\!M}^\perp
  =2V_{M\!M}^\times
   \nonumber \\
   &&\!\!\!\!\!\!\!\!\!\!\!\!\!\!\!\!\!\!\qquad
  -\frac{(t_{M\!M}^\perp)^2}
        {4\beta_M^2/K_{M\!X}^\parallel-U_M
        +4(V_{M\!M}^\parallel-V_{M\!M}^\times)+V_{M\!M}^\perp},
   \label{E:+++vs-++}
\end{eqnarray}
provided $\sqrt{t_{M\!M}^\perp K_{M\!X}^\parallel}\ll\beta_M$, and it is
indeed in excellent agreement with the numerical calculation.
The standard parametrization of $V_{M\!M}^\perp=\sqrt{2}V_{M\!M}^\times$
applied to decoupled chains results in a $(+++)$-$M$-CDW ground state.
It is the advantage of interchain electron hopping that enables
$({\mbox{\large \boldmath$-$}}++)$-$M$-CDW to replace $(+++)$-$M$-CDW.
The close competition between $(+++)$-$M$-CDW and
$({\mbox{\large \boldmath$-$}}++)$-$M$-CDW depends on electron-lattice
couplings as well.
With increasing $\beta_M$,
$({\mbox{\large \boldmath$-$}}++)$-$M$-CDW loses the advantage of
interchain electron transfer $\propto(t_{M\!M}^\perp)^2$ and is again
replaced by $(+++)$-$M$-CDW, as is demonstrated in Fig. \ref{F:PhD}(b).
In the moderate coupling region
$({\mbox{\large \boldmath$-$}}++)$-$M$-CDW generally plays the ground
state, while with increasing electron-phonon interactions of the
Peierls and/or Holstein types, $(+++)$-$M$-CDW appears instead.
\begin{figure*}
\centering
\includegraphics[width=160mm]{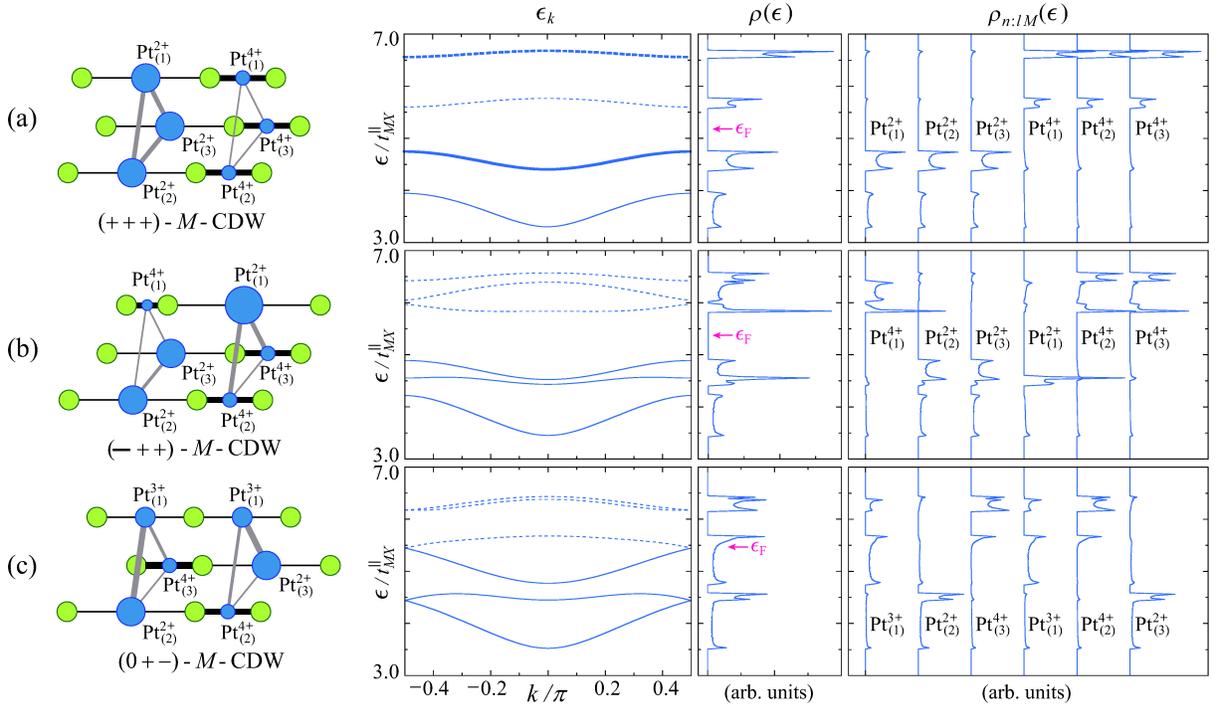}
\vspace*{-2mm}
\caption{(Color online)
         The single-particle energy dispersion relation $\epsilon_k$ and
         the local (total) density of states $\rho_{n:lA}(\epsilon)$
         [$\rho(\epsilon)\equiv\sum_{n,l,A}\rho_{n:lA}(\epsilon)$] for
         $(+++)$-$M$-CDW (a),
         $({\mbox{\large \boldmath$-$}}++)$-$M$-CDW (b) and
         $(0+-)$-$M$-CDW (c).
         We focus on the highest-lying six bands mainly of Pt character,
         where solid and dotted lines consist of occupied and vacant
         states, while thin and thick lines denote singly and doubly
         degenerate bands, respectively.
         $\epsilon_{\rm F}$ indicates the Fermi level.}
\label{F:rho}
\end{figure*}

\begin{figure}
\centering
\includegraphics[width=85mm]{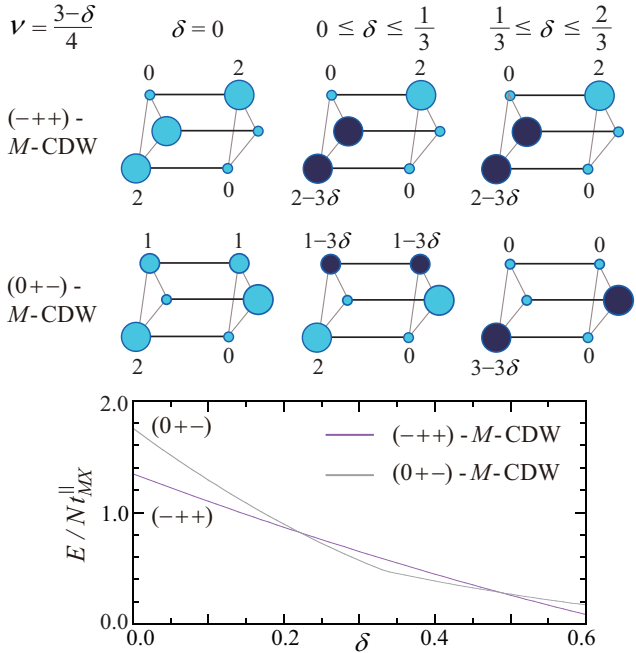}
\vspace*{-2mm}
\caption{(Color online)
         Electron occupancy of the ${\rm Pt}\,d_{z^2}$ orbitals as a
         function of the density of doped holes under strong valence
         localization on the assumption that the $X$ $p_z$ orbitals remain
         fully filled and inactive.
         Energy estimates (\ref{E:Eh(-++)})-(\ref{E:Eh(0+-)2}) are
         plotted as functions of $\delta\equiv 3-4\nu$, where
         $\varepsilon_M$ is set equal to zero.}
\label{F:doping}
\end{figure}

   When we tune the band filling, we find $M$-CDW transitions between
the $(+++)$ and $({\mbox{\large \boldmath$+$}}++)$ types as well as those
between the $(+++)$ and $({\mbox{\large \boldmath$-$}}++)$ types in the
electron-doped region [Fig. \ref{F:PhD}(c)].
The collapse of the electron-hole symmetry results from the
frustration-induced interchain charge polarization.
Although
$({\mbox{\large \boldmath$+$}}++)$-$M$-CDW and
$({\mbox{\large \boldmath$-$}}++)$-$M$-CDW are born of the same
representation X$E'(1)\otimes\check{S}^{0}\otimes\check{T}^{0}$,
their boundaries to $(+++)$-$M$-CDW are different in character.
The invariance groups of $({\mbox{\large \boldmath$\pm$}}++)$-$M$-CDW and
$(+++)$-$M$-CDW are $\mathbf{C}_{2v}\mathbf{L}_{2}\mathbf{ST}$ and
$\mathbf{D}_{3h}\mathbf{L}_{2}\mathbf{ST}$, respectively.
Since the former is a subgroup of the latter, there may be a continuous
transition of the second order between them, which is the case with
any instability arising from PM of the full symmetry.
$(+++)$-to-$({\mbox{\large \boldmath$+$}}++)$-$M$-CDW transitions are
indeed of the second order, whereas
$(+++)$-to-$({\mbox{\large \boldmath$-$}}++)$-$M$-CDW transitions are
of the first order.
A continuous transition between distinct CDW states is unusual and has
never been observed in any other $M\!X$ compound.
We thus wait eagerly for geometrically frustrated $M\!X$ prisms to be
fabricated.

   Figure \ref{F:PhD}(c) stimulates another interest in platinum-halide
nanotubes.
There appears a quite interesting phase, $(0+-)$-$M$-CDW, in the
hole-doped region, which consists of Peierls-insulating dimerized chains
and a paramagnetic regular chain.
Figure \ref{F:rho} elucidates its energy structure, together with those
of $(+++)$-$M$-CDW and $({\mbox{\large \boldmath$-$}}++)$-$M$-CDW.
$(+++)$-$M$-CDW and $({\mbox{\large \boldmath$-$}}++)$-$M$-CDW are fully
gapped, whereas $(0+-)$-$M$-CDW is gapless and metallic.
The metallic band is essentially composed of trivalent platinum ions
[Fig. \ref{F:rho}(c)], though there survives a finite contribution from
those of trapped valence under the standard parametrization of
intermediate electron-lattice coupling.
As a rule for $(0+-)$-$M$-CDW, holes are doped into a single chain and the
rest two remain half filled until the metallic chain is emptied.
In $({\mbox{\large \boldmath$-$}}++)$-$M$-CDW, on the other hand,
the highest-lying filled band is made from a couple of Peierls-distorted
chains in phase [Fig. \ref{F:rho}(b)] and therefore, doped holes
immediately eat into the two CDW chains.
The two Peierls-distorted chains in $(0+-)$-$M$-CDW are free from doped
holes and thus remain stable, while two of the Peierls-distorted chains in
$({\mbox{\large \boldmath$-$}}++)$-$M$-CDW suffer from hole doping and are
thus destabilized.
Once the conduction electrons are removed out, $(0+-)$-$M$-CDW is no more
free from holes invading divalent platinum ions and suffers a rapid
collapse.
Thus ans thus, there may be a re-entrant transition between
$({\mbox{\large \boldmath$-$}}++)$-$M$-CDW and $(0+-)$-$M$-CDW.
Figure \ref{F:doping} illustrates hole injection into $(-++)$-$M$-CDW and
$(0+-)$-$M$-CDW under strong valence localization.
Their $d$-electron energies are evaluated as
\begin{eqnarray}
   &&\!\!\!\!
   \frac{E_{(-++)\mbox{-}M\mbox{-}{\rm CDW}}}{N}
  =3(1-\delta)\varepsilon_M
  +\frac{2+(2-3\delta)^2}{4}U_M
   \nonumber \\
   &&\!\!\!\!\qquad
  +\frac{(2-3\delta)^2}{2}V_{M\!M}^\perp
  +4(2-3\delta)V_{M\!M}^\times
   \nonumber \\
   &&\!\!\!\!\qquad
  -\frac{2+(2-3\delta)^2}{2}\frac{\beta_M^2}{K_{M\!X}^\parallel}
   \ \ \ \ \left(0\leq\delta\leq\frac{2}{3}\right),
   \label{E:Eh(-++)}
   \\
   &&\!\!\!\!
   \frac{E_{(0+-)\mbox{-}M\mbox{-}{\rm CDW}}}{N}
  =3(1-\delta)\varepsilon_M
  +\frac{4+(1-3\delta)^2}{4}U_M
   \nonumber \\
   &&\!\!\!\!\qquad
  +(1-3\delta)^2V_{M\!M}^\parallel
  +2(1-3\delta)V_{M\!M}^\perp
  +4(2-3\delta)V_{M\!M}^\times
   \nonumber \\
   &&\!\!\!\!\qquad
  -\frac{2\beta_M^2}{K_{M\!X}^\parallel}
   \ \ \ \ \left(0\leq\delta\leq\frac{1}{3}\right),
   \label{E:Eh(0+-)1}
   \\
   &&\!\!\!\!
   \frac{E_{(0+-)\mbox{-}M\mbox{-}{\rm CDW}}}{N}
  =3(1-\delta)\varepsilon_M
  +\frac{9(1-\delta)^2}{4}U_M
   \nonumber \\
   &&\!\!\!\!\qquad
  +9(1-\delta)^2V_{M\!M}^\times
   \nonumber \\
   &&\!\!\!\!\qquad
  -\frac{9(1-\delta)^2}{2}\frac{\beta_M^2}{K_{M\!X}^\parallel}
   \ \ \ \ \left(\frac{1}{3}\leq\delta\leq\frac{2}{3}\right),
   \label{E:Eh(0+-)2}
\end{eqnarray}
and are plotted in Fig. \ref{F:doping}.
In $(0+-)$-$M$-CDW doped holes are distributed differently according as
their density, defined as $3-4\nu\equiv\delta$, exceeds one third or not.
That is why $(0+-)$-$M$-CDW once replaces $(-++)$-$M$-CDW but soon
disappears with decreasing electron occupancy.
The naivest estimates (\ref{E:Eh(-++)})-(\ref{E:Eh(0+-)2}) succeed in
predicting such a re-entrant transition, though they overstabilize
$(0+-)$-$M$-CDW against $(-++)$-$M$-CDW.
The preceding transition point is well explained, but the succeeding one
is harder to understand.
It is likely that linear corrections of hopping energy to both
$E_{(-++)\mbox{-}M\mbox{-}{\rm CDW}}$ and
$E_{(0+-)\mbox{-}M\mbox{-}{\rm CDW}}$ should make their competition under
doping much less intuitive.
\begin{figure*}
\centering
\includegraphics[width=160mm]{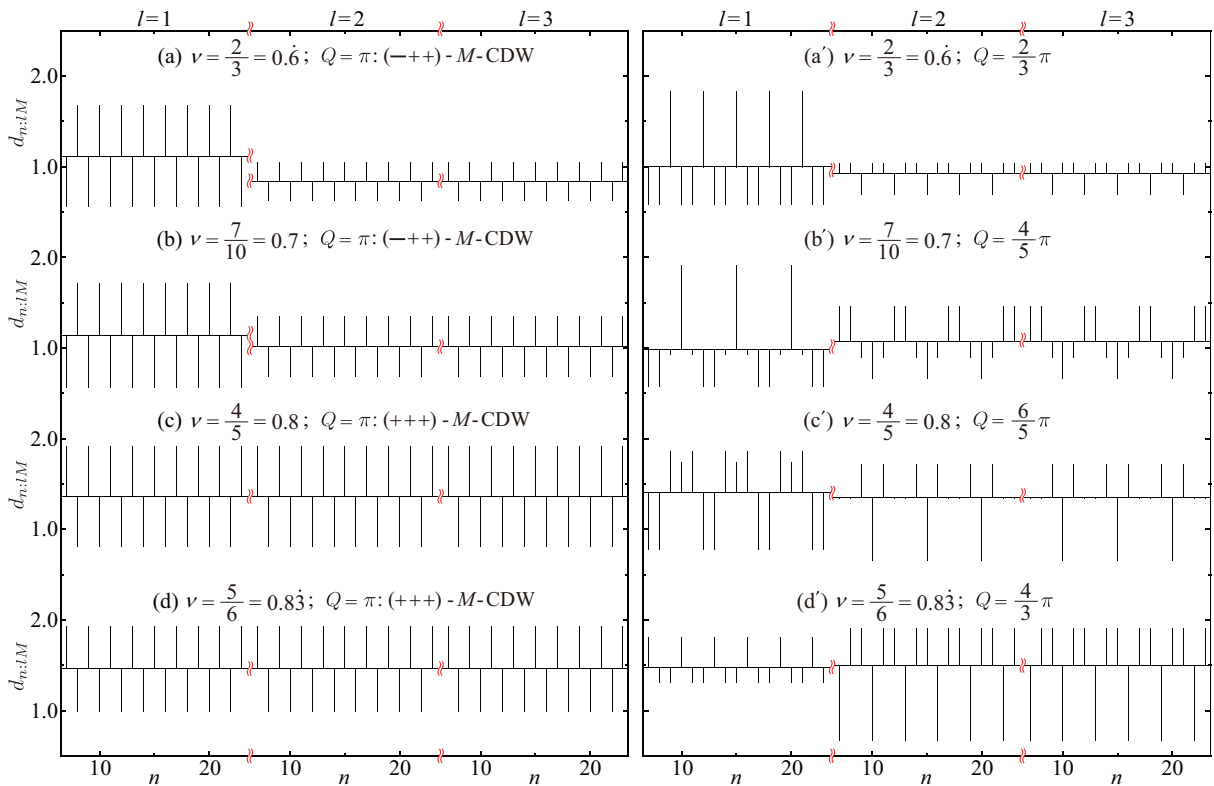}
\vspace*{-2mm}
\caption{(Color online)
         Metastable CDW states of $Q\neq\pi$ against the ground states,
         with $Q=\pi$, at various values of $\nu$, where the local
         electron densities $d_{n:lM}$ are measured in comparison with the
         average occupancy of the relevant chain labeled $l$.
         Seventeen metal triangles are clipped out of the prism of
         $N=300$.}
\label{F:LPCDW}
\end{figure*}

\section{Summary and Discussion}

   Platinum-halide triangular prisms are thus identified as frustrated
Peierls-Hubbard nanotubes and are productive of novel CDW states.
Any structural instability is conditional and the homogeneous CDW state
of $\mathbf{D}_{3h}$ symmetry is broken down into those of lower symmetry
with increasing interchain transfer energy and/or varying electron
occupancy.
The geometric frustration unbalances three chains in charge and yields
heterogeneous CDW states, which are possibly detectable by NMR
chemical-shift \cite{K40} and X-ray scattering \cite{W6676} measurements.

   Electrochemical doping possibly causes successive phase transitions
between totally valence-trapped and fully metallic states.
We encounter both first- and second-order insulator-to-metal transitions
under hole doping, whereas we find continuous transitions between distinct
CDW states under electron doping.
A stepwise increase in conductivity is expected of the discontinuous
insulator-to-metal transition.
It is the frustration-induced interchain charge polarization rather than
the two-band modeling employed that breaks the electron-hole symmetry.

   There is a possibility of structural instabilities of longer period
existing under doping.
Indeed we have extensively investigated solutions of $Q\neq\pi$, including
incommensurate phases, but none of them competes in energy with those of
$Q=\pi$ in the present modeling.
Some examples are shown in Fig. \ref{F:LPCDW}.
At $\nu=2/3$ and $5/6$ the ${\rm Pt}\,d_{z^2}$ orbitals are effectively
$1/3$- and $2/3$-filled, where formally commensurate, that is to say,
cell-tripled, $M$-CDW states are found, while at $\nu=7/10$ and $4/5$ the
${\rm Pt}\,d_{z^2}$ orbitals are effectively $2/5$- and $3/5$-filled,
where subharmonic, namely, cell-quintupled, $M$-CDW states are found.
Trimerized CDW solutions are available at
$0.646\alt\nu\alt 0.694$ and $0.805\alt\nu\alt 0.847$, whereas
pentamerized ones are detectable at
$0.691\alt\nu\alt 0.733$ and $0.783\alt\nu\alt 0.833$, but they are never
stabilized into the ground state under the standard parametrization.
With increasing $\beta$, however, they closely compete in energy with
dimerized CDW solutions and finally become preferable in general.
In the cases of $\nu=2/3$ and $5/6$, the transition points are estimated
at
$\beta/\sqrt{t_{M\!X}^\parallel K_{M\!X}^\parallel}\simeq 0.705$ and
$0.814$, respectively,
while in the cases of $\nu=7/10$ and $4/5$, they turn out
$\beta/\sqrt{t_{M\!X}^\parallel K_{M\!X}^\parallel}\simeq 0.765$ and
$0.855$, respectively, all of which may be too large to be realized in
available platinum halides.
It was predicted for a single $M\!X$ chain that incommensurate ground
states should appear even at $3/4$ filling provided the site-diagonal
electron-lattice coupling is sufficiently strong, \cite{B13228} but they
are not yet observed experimentally.

   There is an attempt at fabricating a platinum-bromide triple-chain
compound. \cite{Ocomm}
Substitution of platinum ions with nickel ions should lead to stronger
correlations between $d$ electrons and might result in novel spin
arrangements.
We hope our calculations will stimulate further chemical explorations of
tubed $M\!X$ compounds.
\vspace{3mm}

\acknowledgments
\vspace{-2mm}
   We are grateful to K. Otubo and H. Kitagawa for valuable information on
their extensive chemical explorations.
This work was supported by the Ministry of Education, Culture, Sports,
Science and Technology of Japan.

\end{document}